# CYCLINACS: FAST-CYCLING ACCELERATORS FOR HADRONTHERAPY

U. Amaldi [a,1], S. Braccini [a,2,*], A. Citterio [a,3], K. Crandall [a], M. Crescenti [a,4], M. Dominietto [a,5], A. Giuliacci [a,6], G. Magrin [a], C. Mellace [a], P. Pearce [a], G. Pittà [a], E. Rosso [a], M. Weiss [a,†] and R. Zennaro [a,7]

[a] *TERA Foundation, Via Puccini 11, Novara, Italy*

**Abstract**

We propose an innovative fast-cycling accelerator complex conceived and designed to exploit at best the properties of accelerated ion beams for hadrontherapy. A cyclinac is composed by a cyclotron, which can be used also for other valuable medical and research purposes, followed by a high gradient linear accelerator capable to produce ion beams optimized for the irradiation of solid tumours with the most modern techniques. The properties of cyclinacs together with design studies for protons and carbon ions are presented and the advantages in facing the challenges of hadrontherapy are discussed.

**In memory of Mario Weiss who lead the developments of cyclinacs**

**from 1993 to 2003**

[*] Corresponding author. Address: University of Bern, Laboratory for High Energy Physics, Sidlerstrasse 5, CH-3012 Bern, Switzerland. E-mail address: Saverio.Braccini@cern.ch (S. Braccini).
[1] Also at University of Milano-Bicocca, Piazza della Scienza 3, 20126 Milano, Italy.
[2] Now at University of Bern, Laboratory for High Energy Physics, Sidlerstrasse 5, CH-3012 Bern, Switzerland.
[3] Now at Paul Scherrer Institut, 5232 Villigen PSI, Switzerland.
[4] Now at European Patent Office, Patentlaan 2, 2288 EE Rijswijk, The Netherlands.
[5] Now at ETH, Institut für Biomedizinische Technik, Wolfgang-Pauli-Str. 10, 8093 Zürich, Switzerland.
[6] Now at IBA Dosimetry GmbH, Bahnhofstraße 5, 90592 Schwarzenbruck, Germany.
[7] Now at CERN, CH-1211 Genève 23, Switzerland.
[†] Deceased.

# 1. INTRODUCTION

Hadrontherapy, the technique of tumour radiotherapy which employs beams of the heavy particles made of quarks, called "hadrons", is developing very rapidly. In 2008 turn key protontherapy centres and "dual" centres – featuring both proton and carbon ion beams – are offered by commercial companies. The accelerators for protontherapy are 4-5 metre diameter cyclotrons, both at room temperature and superconducting, and 6-8 metre diameter synchrotrons while for carbon ion therapy only 20-25 metre diameter synchrotrons are in use. Recently a company has started the construction of a large superconducting cyclotron for carbon ions.

In the 60 years elapsed from Wilson's proposal to make good use of the "Bragg peak" [1], more than 55,000 patients have undergone protontherapy and this radiotherapy technique is booming, while 7000 patients have been treated with ions, most of them with carbon ions [2].

Protontherapy will continue to develop and one can expect that in ten years for every 10 million inhabitants about 20,000 patients every year will be continued to be treated with high energy photons ("X rays" in the radiation oncologist parlance) and about 2500 with protons. Also the therapy of radio resistant tumours with carbon ions is rapidly developing. Two carbon ion and proton 'dual' centres are running in Japan, in the Prefectures of Chiba and Hyogo, and two centres are in the commissioning phase in Europe, in Heidelberg and Pavia. Here the centre proposed and designed by TERA [3] is built under the responsibility of the CNAO Foundation [4].

In this paper a novel fast-cycling accelerator is proposed which produces charged hadron beams better adapted to hadrontherapy than the ones which are extracted from cyclotrons and synchrotrons. A "cyclinac" is based on a cyclotron, or a synchrocyclotron, and a high-gradient linac. The first high-frequency hadron linac ever built is described in Section 4, after a discussion of the challenges faced by hadrontherapy and a description of how active dose delivery systems can be used to respond to the challenges (Section 2). The "cyclinac' concept and its first applications are described in Section 3 and 5. More detailed examples of cyclinacs designs for protontherapy and carbon ion therapy are given in Sections 6, 7 and 8.

# 2. THE CHALLENGE CONFRONTING HADRONTHERAPY

The numbers given above are impressive. However, about 75% of all the patients treated with protontherapy have been irradiated in nuclear and particle physics laboratories by means of non-dedicated accelerators. Moreover only 1% of all these patients have been treated with 'active' delivery systems in which the tumour target is uniformly 'painted' with a large number of successive 'spots' thus making the best possible use of the properties of charged hadron beams. This fundamental technical advance took place at the end of the century in two physics laboratories: the *Paul Scherrer Institute* (PSI in Villigen, Switzerland) where the 'spot scanning' technique was developed for protons [5], and the *Gesellschaft für Schwerionenforschung* (GSI in Darmstadt, Germany) where the 'raster scanning' technique was developed for carbon ions [6]. Still in 2008 almost all hospital-based centres are still using 'passive' dose delivery systems in which the beam is scattered in successive targets and flattened and/or shaped with appropriate filters and collimators [7]. In some centres, the more advanced semi-active 'layer stacking' technique is used [8].

In the next years hadrontherapy centres have to implement new approaches for the delivery of the dose if they want to keep pace with the competition of conventional radiotherapy – performed mainly with X rays produced by electron linacs. Indeed new techniques have been introduced in the last ten years to 'conformally' cover with many crossed beams moving tumours and spare more and more the surrounding healthy tissues. Many hospitals apply routinely *Intensity Modulated Radiation Therapy* (IMRT) [9] and start to use *Image Guided Radiation Therapy* (IGRT) [10] and *Tomotherapy* [11]. Hadron dose delivery systems have to become more sophisticated in order to bring to full fruition the intrinsic advantages of the dose distribution due to a single narrow ion beam characterized by the well-known 'Bragg peak'.

Proton beams of energy between 200 and 250 MeV (and very low currents, of about 2 nA) and carbon ion beams of energy between 3500 and 4500 MeV (and currents of about 0.2 nA) are advantageous in the treatment of deep-seated tumours because of four physical properties [12]. Firstly, they deposit their maximum energy density abruptly at the end of their range. Secondly, they penetrate the patient practically without diffusion (concerning this property carbon ion beams are from three to four times better than proton beams). Thirdly, being charged, they can easily be formed as narrow focused and scanned pencil beams of variable penetration depth so that any part of a tumour can be accurately irradiated. The fourth physical property is linked to radiobiology and pertains to ions, carbon ions in particular: since each ion leaves in a traversed cell about 24 times more energy than a proton having the same range, the damages produced in crossing the DNA of a cell nucleus are different since they include a large proportion of multiple close-by double strand breaks. These damages cannot be repaired by the usual cellular mechanisms so that the effects are *qualitatively* different from the ones produced by the other radiations and carbon ions can control tumours which are radio resistant to both protons and X rays [13].

The first property is the main reason for using charged hadrons in radiotherapy since the single beam dose distribution is in all cases superior to the one of X-rays, which has an almost exponential energy deposition in matter after a maximum dose delivered only few centimetres inside the patient's body. Beams of charged hadrons allow by principle a more conformal treatment of deep-seated tumours than beams of X rays; they give minimal doses to the surrounding tissues, and - in the case of carbon ions - open the way to the control of radio resistant tumours.

The challenge of hadrontherapy is in making full use of the above four properties especially when the tumour moves, for instance because of the breathing of the patient. The fact that protons and ions have an electric charge, the third property, is the key to any further development but, surprisingly enough, till now practically all therapy beams have been 'formed' by collimators and absorbers as if hadrons had no electric charge.

In the GSI active 'raster scanning' technique, a pencil beam of 4-10 mm width (FWHM) is moved in the transverse plane almost continuously by two bending magnets along a trajectory similar to the one of the electron beam which scans a TV screen. After painting a section of the tumour, the energy of the beam extracted from the carbon ion synchrotron is reduced to paint a less deep layer. In practice to obtain a variable speed the beam is moved in steps much smaller than the FWHM of the spot and the next small step is triggered when a predetermined integral of the fluency has been recorded by the ionization chambers placed upstream of the patient. In such an approach the beam is always 'on'.

In the PSI active 'spot scanning' technique, the 8-10 mm (FWHM) spot is moved by much larger steps (of the order of the FWHM of the spot) and, as in the previous case, the transverse movement – which takes about 2 ms - is triggered by ionization chambers measuring the fluence. During the movement of the spot the proton beam extracted from the cyclotron is interrupted for 5 ms by means of a fast kicker.

In both cases the tumour target is 'painted' only once and this is an inconvenience in the case of moving organs since any movement can cause important local under-dosages or over-dosages. Three strategies have been considered to reduce such effects. In order of increasing complexity they are:

(i) in the irradiation of the thorax and the abdominal region the dose delivery is synchronized with the patient expiration phase in a process called "respiratory gating" so that the effects on the distribution of the dose due to the movements of the organs are reduced to a minimum (this technique is already used also in conventional radiotherapy);

(ii) the tumour is painted many times in three dimensions so that the movements of the organs (if not too large) can cause only small (≤3%) over- dosages and/or under-dosages;

(iii) the movement is detected by a suitable system, which outputs in real-time the 3D position of the tumour, and a set of feedback loops compensate for the predicted position in the dose delivery plan with on-line adjustments of the transverse and longitudinal locations of the following spots, as shown in Fig.1 [14].

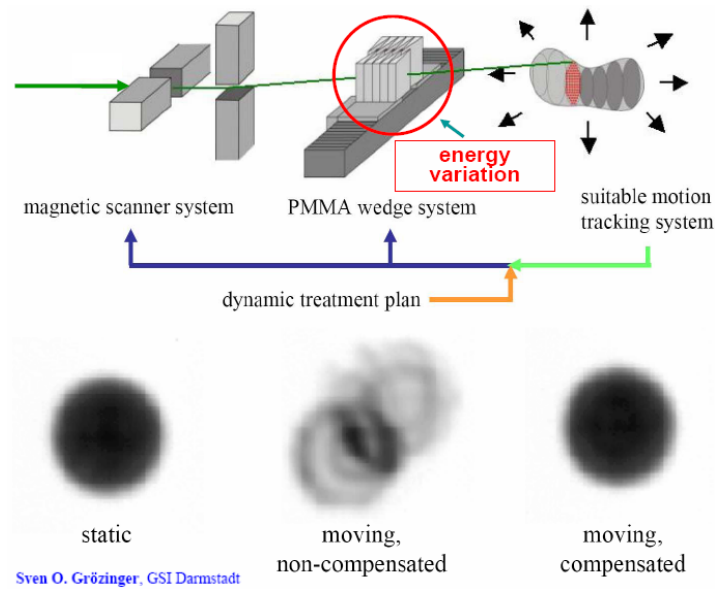

*Figure 1. The feedback system - studied numerically and experimentally at GSI - compensates for the movements of the organs acting, with two bending magnets, to correct the transverse movements and, with absorbers of variable thickness, to compensate for longitudinal movements.*

An optimal delivery mechanism should be such to allow the use of any combination of these three approaches: respiratory gating, multi-painting and active angular/energy feedbacks.

## 3. CYCLINACS AND THEIR PROPERTIES

From 1993, and in parallel with the work done for CNAO, TERA has proposed [15] and developed a novel type of accelerator – the "cyclinac" - which produces charged hadron beams which respond to the above requests better than cyclotrons and synchrotrons.

The first proposal concerned a 30 MeV cyclotron and a proton linac running at the same 3 GHz frequency used by the electron linacs employed in conventional radiotherapy (Fig. 2). This would imply high gradients and thus a relativly short accelerator.

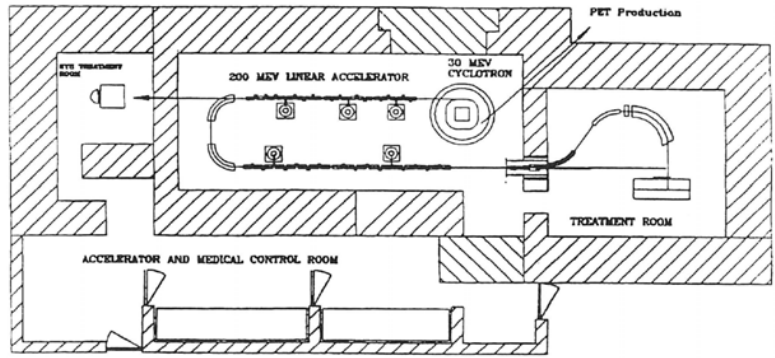

*Figure 2. The first sketch of what was later called a "cyclinac" was based on a 30 MeV commercial cyclotron used also for the production of radiopharmaceuticals [16].*

The study soon branched out in two approaches described in the "Green Book" [16].

Firstly, due to the stringent mechanical requirements needed to construct a linac capable of accelerating 30 MeV protons ($\beta = 0.25$), it was decided that the first Side Coupled Linac (SCL) would be designed for a 62 MeV input energy, having in mind in particular the Clatterbridge cyclotron. The results of the optimization were presented by M. Weiss and K. Crandall in 1994 [17]. Since then M. Weiss has been the leader of the project LIBO (LInac BOoster) prototype. Shortly afterwards R. Zennaro joined as responsible for the design and the RF tests of the accelerating structures.

Secondly, an "all-linac solution" was studied by L. Picardi et al., which is based on a Side Coupled Drift Tube Linac having an input energy below 62 MeV followed by a LIBO structure for higher energies. This is the "TOP project" of ENEA and ISS [18].

In 1998 a collaboration was set up among TERA, CERN (E. Rosso, B. Szelezs et al.), the University and INFN of Milan (C. De Martinis et al.) and the University and INFN of Naples (V. Vaccaro et al.) to build and test the double-module LIBO prototype which is at present on display at the CERN "Physics and Health" exhibition. This item is discussed in Section 4.

A generic cyclinac is shown in Fig. 3.

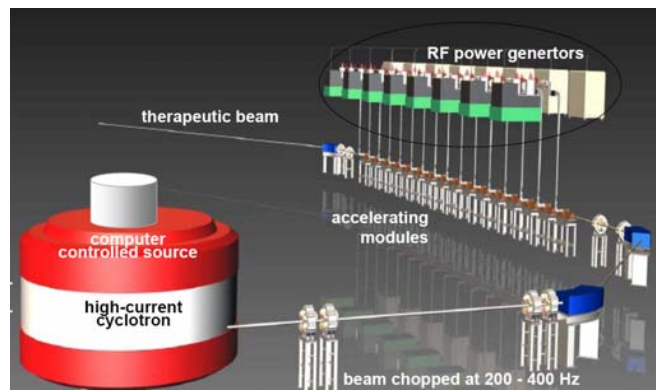

*Figure 3. Sketch of a cyclinac with the cyclotron and the high gradient linac. One beam is sent to the linac. Other beams extracted from the cyclotron can be used for medical purposes different from hadrontherapy.*

The hardware components of a cyclinac are:
  (i)   a computer controlled ion source,
  (ii)  a cyclotron (or a synchrocyclotron or, in the future possibly a Fixed Field Alternating Gradient – FFAG - accelerator),
  (iii) some external beams typically used for other purposes than the therapy allowed by the linac,
  (iv)  a beam transport system which brings the particles extracted from the cyclotron to the linac,
  (v)   a linac, which accelerates the cyclotron beam to the required energy for therapy treatments is generally composed of a Drift Tube Linac section and a Side Coupled Linac section,
  (vi)  a distribution system of the high-energy beam to the treatment rooms, equipped with rotating gantries or fixed beams.

It is important to underline that the indicated "other beams" extracted from the cyclotron are used for different valuable medical, non medical and research applications such as radioisotope production for diagnostics and endoradiotherapy.

The source is triggered at the repetition rate of the linac and the beam is chopped accordingly, so that a very low current is sent to the linac gallery. This repetition rate (as shown in Fig. 3) is chosen having in mind the spot scanning technique developed by PSI [5] in which, to 'paint' the tumour, the Bragg 'spot' is moved transversally with two crossed magnetic fields.

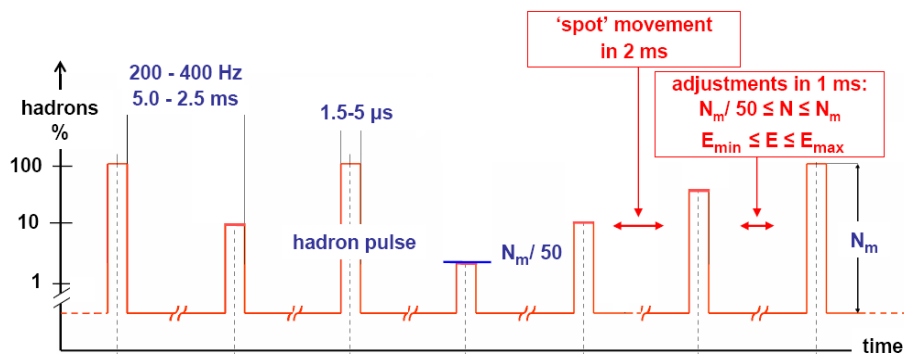

*Figure 4. Time and amplitude structures of the therapeutic beam produced by a cyclinac.*

As presented in Fig. 4, the linac produces 1.5-5 µs long "hadron pulses" separated by 5 ms (2.5 ms) when the repetition rate of the klystrons producing the RF power is 200 Hz (400 Hz). Note that the pulsed beam is continuously present, as in a cyclotron.

In the 2.5-5 ms separating two successive pulses, the number of particles to be accelerated in the next "hadron pulse" and to be delivered to the next voxel can be adjusted with a 3% precision - between $N_m$ and $N_m/50$ - by acting on the computer controlled source of the cyclotron. In parallel the hadron energy can be chosen by acting on the amplitudes of the RF pulses powering the twenty or so accelerating modules. To this end, the number of accelerating cells per module and the strengths of the Permanent Magnetic Quadrupoles (PMQs) – two per module – are chosen in such a way that a beam of any energy is transported without losses along the linac. To reduce the energy, the RF power sent to a contiguous set of accelerating modules – starting from the end of the linac - is set to zero while the power sent to the last active one is continuously varied. Beam dynamics computing programs by K. Crandall were used to define the optimal conditions.

As indicated in the last row of Table 1, the beam produced by a linear accelerator is more flexible and its intensity and time structure fit the requirements of spot scanning and allow an easy implementation of the three strategies for treating moving organs described at the end of the last Section. This is mainly because in linear accelerators there is no need of complex injection and extraction systems. It is important to remark that the energy can be varied in 1 ms only.

*Table 1. Properties of the beams of various accelerators.*

| Accelerator | The beam is always present? | The energy is electronically adjusted? | What is the time to vary $E_{max}$? |
|---|---|---|---|
| Cyclotrons | Yes | No | ≥50 ms |
| Synchrotrons | No | Yes | 1 s |
| Cyclinacs | Yes | Yes | 1 ms |

To understand the basic arguments behind these statements, it has to be recalled that, as shown in Fig. 4, a linear accelerator produces 1.5-5 µs long hadron pulses separated by several milliseconds, typically 5 ms (2.5 ms) when the repetition rate of the klystrons producing the RF power is 200 Hz (400 Hz).

The *first question* of Table 1 is relevant when the beams are synchronized with the expiration phase of the irradiated patient. The *second question* is important for the depth scanning of tumours. Cyclotrons require the mechanical movement of absorbers to vary the penetration range, while the electronic control, possible in synchrotrons and cyclinacs, is more reliable, requires less maintenance and does not require an additional Energy Selection System (ESS) that produces neutrons and induced radioactivity. However, in synchrotrons the adjustment takes one second and cannot be used for the 3D feedback system proposed by GSI to treat moving organs and tested by varying the depth with fast moving absorbers [14]. Instead, in a cyclinac the variation of the depth of the Bragg 'spot' is even faster that the transverse adjustment. To profit from this feature to treat longitudinally moving organs it is enough for the transport line and the gantry to have a ± (1.5 - 2)% momentum acceptance, so that the depth can be varied in about one millisecond by ± (10 - 15) mm.

A further interesting property of a cyclinac is connected to the in-beam-PET devices used during patient irradiation. Since many years at GSI the dose delivered by the carbon ion-beam is controlled by measuring during treatment the distribution of the positron emitting radioisotopes produced in the irradiated tissues by means of a PET scanner located around the patient's body. More recently, studies have been performed to apply the same technique to the localization of the dose deposition in the case of proton irradiations [19]. The activity is large enough and the distal fall-off is so sharp to allow a range determination with 1 mm accuracy. This opens the way to an important quality control which most probably will be applied in future protontherapy and ion therapy centres.

In this respect a linac is ideal because the beam is off for more than 99% of the time. This avoids two difficulties encountered, for instance, with synchrotrons:

(1) the problem caused by the accidental coincidences due to high energy photons produced in the irradiated patient during the extraction of the beam, for which a sophisticated hardware and software have been developed;

(2) the complication due to the wash-out of the induced radioactivity by the blood flow, which in principle can be corrected for, but depends upon unknown parameters.

It has to be remarked that using a carbon ion cyclinac the in-beam PET detector will also collect the coincidence signals due to short-lived positron emitting isotopes: $^8$B ($T_{1/2}$ = 770 ms), $^9$C (127 ms), $^{12}$N (11 ms), $^{13}$O (8.6 ms) and $^{14}$O (70.5 ms) with a (small) increase of the PET statistics and negligible deterioration of the spatial accuracy.

## 4. DEVELOPMENTS AND TESTS OF THE LIBO PROTOTYPE

The fraction of a continuous beam transmitted by a linac with the time structure of Fig. 4 is in the range $10^{-5}$–$10^{-4}$. The main point of this paper is that, in the case of hadrontherapy, such a minute overall acceptance does not pose any problem because – as remarked above – tumour therapy with protons and carbon ion beams requires beam currents of only 2 nA or 0.2 nA, respectively. These very small currents are easily obtained if the linac is placed downstream of a commercial cyclotron capable of producing without problems $10^6$–$10^7$ times larger currents. This fact has the added advantage that, if so desired, these high currents can in parallel produce radiopharmaceuticals for diagnostics, pain palliation and tumour therapy or be used for research purposes.

Since 1993 the linac frequency (f = 3 GHz) was chosen to be equal to the one used by the about 10,000 linacs used in the world to accelerate electrons for conventional X-ray therapy. It is a high - never used before - frequency for accelerating hadrons but it allows a shorter linac design since the accelerating gradient is roughly proportional to $[f]^{1/2}$.

As mentioned above, the initial proposal concerned a 30 MeV cyclotron, but, when the prototyping was initiated under the leadership of M. Weiss, it was decided that the first linac prototype would be designed for a 62 MeV input energy, having in mind in particular the Clatterbridge cyclotron used for the therapy of eye melanomas as a possible injector [16]. In 1998 a collaboration was set-up among TERA, CERN, the University and INFN of Milan and the University and INFN of Naples which built and tested the first high-frequency proton linac which was dubbed LIBO (LInac BOoster).

The LIBO prototype is a 3 GHz Coupled Cell Linac (CCL) with a design gradient of 15.7 MV/m. As shown in Fig. 5, the prototype is composed of four accelerating 'tanks' each made of 23 half-cell-plates braised together. The module, 1.3 m long, is powered through a single central bridge coupler connected to a klystron. During the power tests, performed in the LIL tunnel at CERN, the design gradient was easily reached by injecting the nominal peak power of 4 MW. With the maximum available power from the klystron a gradient up to 27 MV/m was reached without discharges [20].

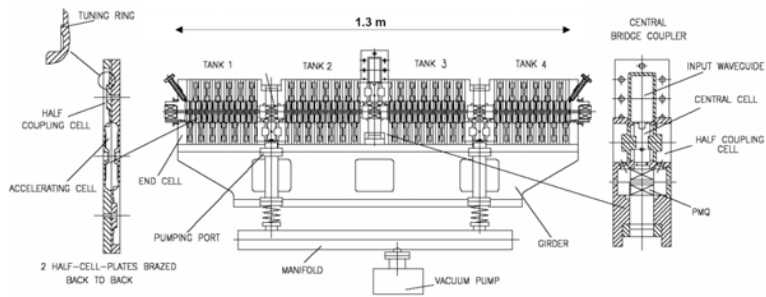

*Figure 5. Mechanical design of the four 'tanks' of the LIBO prototype. Each tank is made of a number of basic units machined with high accuracy in copper and called 'half-cell-plates'. Permanent Magnetic Quadrupoles (PMQ) are located between two successive tanks to focus the accelerated proton beam.*

After the final tests of the prototype, performed at the *Laboratori Nazionali del Sud* of INFN in Catania, where protons were accelerated as predicted [21], it was possible to reconsider the initial idea of a 3 GHz proton linac starting from 30 MeV. By decreasing the cyclotron energy from 62 MeV to 30 MeV the proton velocity reduces by √2 and this implies thinner half-cells that are more difficult to produce and braise. The technical problems have been solved and an accelerating module made of two tanks has been built, as presented in Fig. 6.

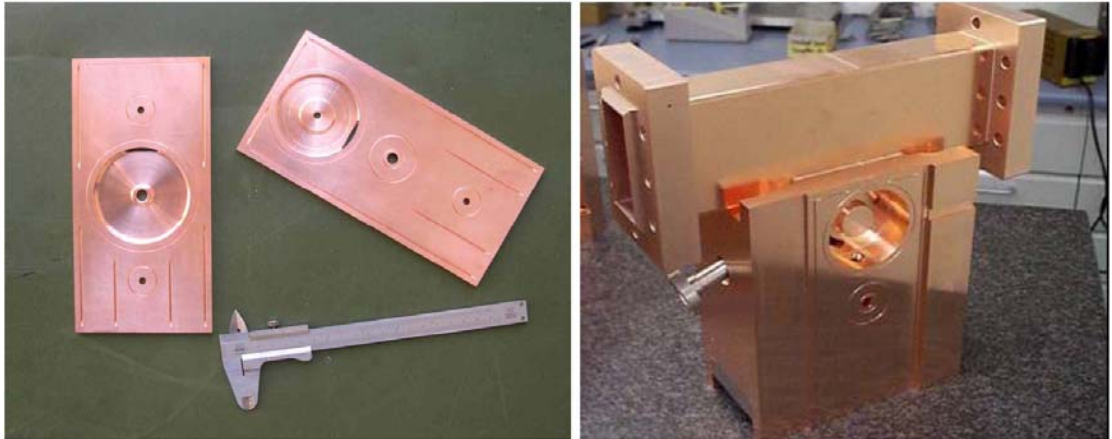

*Figure 6. Two half-cells (left figure) and the bridge coupler of the 50 cm long module - made of two tanks- which accelerates protons from 30 MeV to 35 MeV (right figure).*

## 5. THE "CYCLINAC": FIRST EXAMPLES

This new type of accelerator implies a low duty cycle of the therapy beam (typically 0.1%) and a small longitudinal acceptance (typically 10%). However cyclotron currents are large (10-500 μA), as mentioned in the last Section, and a cyclinac can easily accelerate currents of the order of 2 nA (0.2 nA) which are needed for therapy with proton (carbon ion) beams.

In the last years TERA has designed various cyclinacs. As examples, Table 2 contains the main parameters of two of them. Details and other solutions are presented in the next Sections. In both schemes the particles are accelerated by a linac which is of a single type which consists of a LIBO structure running at 3 GHz.

The first scheme is based on a cyclotron which accelerates protons up to 30 MeV. Two companies offer high-current turn-key cyclotrons of the needed properties: IBA (Ion Beam Applications, Belgium) and ACSI (Advanced Cyclotron System Inc., Canada). These machines are such that currents as large as 750-1500 μA can be extracted.

The second design is based on a novel superconducting cyclotron that accelerates carbon ions $C^{+6}$ to 250 MeV/u. This cyclotron has been designed by L. Calabretta et al. of *Laboratori Nazionali del Sud* of INFN in Catania [22] . More recently the Catania group has increased the energy of this cyclotron from 250 MeV/u to 300 MeV/u. Of course, such a new situation is easily handled by suppressing the initial accelerating modules which accelerate ions from 250 MeV/u to 300 MeV/u.

This cyclotron has been dubbed SCENT (*Superconducting Cyclotron for Exotic Nuclei and Therapy*) and accelerates $H_2^{+1}$ hydrogen molecules - which are extracted from the cyclotron in the form of single protons by stripping in a thin foil - and also carbon ions $C^{+6}$, extracted through the same magnetic channel by a deflector. The 250 MeV protons are used for protontherapy. The 3000 MeV (250 MeV/u x12) carbon ions penetrate 12.8 cm of water while the output beam of the linac has 4800 MeV (400 MeV/u) and reaches depths up to 27.9 cm of water, allowing the treatment of all deep seated tumours.

**Table 2.** *Main parameters of the proton and the carbon ion linacs (f = 2998 GHz). In the last column the number indicates the row used in the text as a reference.*

| **Accelerated particles** | $p^{+1}$ | $C^{+6}$ | |
|---|---|---|---|
| Type of linac | LIBO | LIBO | R1 |
| Input energy [MeV/u] | 30 | 250 | R2 |
| Output energy [MeV/u] | 236 | 400 | R3 |
| Cells per tank / tanks per module | 14/2 | 15/2 | R4 |
| Number of accelerating modules | 22 | 18 | R5 |
| Diameter of the beam hole [mm] | 5.0 | 8.0 | R6 |
| Total length of the linac [m] | 18.7 | 24.0 | R7 |
| Number of Permanent Magnetic Quadrupoles (PMQ) | 45 | 37 | R8 |
| Length of each PMQ (with gradients 120-170 T/m) [mm] | 30 | 60 | R9 |
| Synchronous phase | -15° | -15° | R10 |
| Peak power per module (with 10% losses) [MW] | 2.6 | 4.2 | R11 |
| Effective shunt impedance $ZT^2$ (inject.-extract.) [M$\Omega$/m] | 22–70 | 81-86 | R12 |
| Average axial electric field (injection-extraction) [MV/m] | 16.4–17.8 | 21.2-20.5 | R13 |
| Total peak RF power for all the klystrons (R5xR11) [MW] | 57 | 76 | R14 |
| Klystron RF efficiency | 0.42 | 0.42 | R15 |
| Peak RF power for all the klystrons (R15:R14) [MW] | 135 | 180 | R16 |

The accelerating modules (R5) are fed by commercial radio-frequency klystrons at present and, maybe, magnetrons in future, if these oscillators could be synchronized to better than 1°. The required high voltage peak-powers are large (R16) but are easily provided by high efficiency modulators in spite of the high frequency because high gradients have been chosen (R13) to keep the linacs reasonably short. About 20 metres are needed (R7) but it is worthwhile noting that such lengths are similar to the ones of the beam transport channels – usually dubbed 'Energy Selection Systems' - that are used to degrade and clean the beam produced by a cyclotron for hadrontherapy.

The operational way to limit the wall plug power is to make the duty cycle as small as possible. Two factors combine in the duty cycle: the repetition rate and the duration of the RF pulses. This issue requires a detailed discussion.

The linac repetition rate is fixed by the requirement of delivering the typical dose for therapy accelerators with a large number of paintings obtained by revisiting many times the same point with the Bragg spot. At PSI and GSI the spot Full Width at Half Maximum (FWHM) is varied in the typical range 4-10 mm. Its choice influences the 80-20% lateral fall-off of the dose which, because of multiple scattering, is *naturally* 6 mm for 230 MeV protons and 2 mm for 400 MeV/u carbon ions. Due to the larger multiple scattering of protons in matter, the FWHM of the spots can be larger than in the irradiation with carbon ions without excessive deterioration of the lateral fall-off.

At PSI the distance between the centres of two spots is equal to 75% of the FWHM. This choice correspond to a ratio (global dose)/(max dose) = 2, i.e. in each point a single spot contributes always by less than half to the total delivered dose. Moreover since the maximum dose non uniformity due to the finite number of spots is ±1.25%, the distance between the spots has to be not larger than 0.75 FWHM.

At a depth of 200 mm in water the 80-20% lateral fall-off due to the multiple scattering of 173 MeV protons (330 MeV/u carbon ions) is 5.46 mm (1.45 mm), which, assuming a Gaussian distribution, corresponds to curve with a FWHM of 11.5 mm (3.1 mm). By combining this natural spread with the finite dimension of a pencil beam having a 4 mm FWHM, the total transverse FWHM values obtained are 12.2 mm and 5.0 mm for protons and carbon ions, respectively.

Fig. 7 shows the normalized number of particles which, placed at the vertex of a 11.7 (4.8) mm lattice with 15.6 (6.3) mm FWHM transverse dimensions, produce a uniform distributions within ±1.25% for all the points of the tumour target which is represeanted by a one litre volume sphere with the centre at a depth of 20 cm. In the proton (carbon) case the lateral 80-20% fall-off of the spot increases from the natural values (p: 7 mm, C: 1.8 mm) to 7.2 mm and 3.0 mm, respectively.

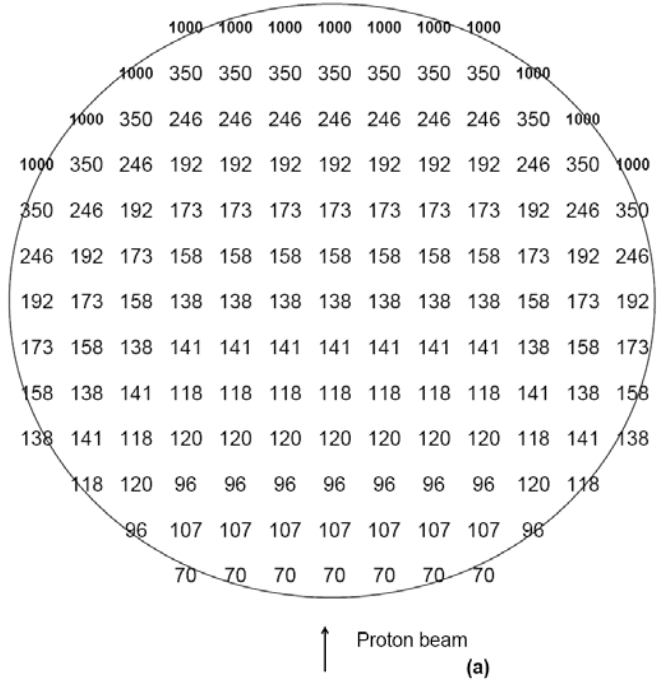

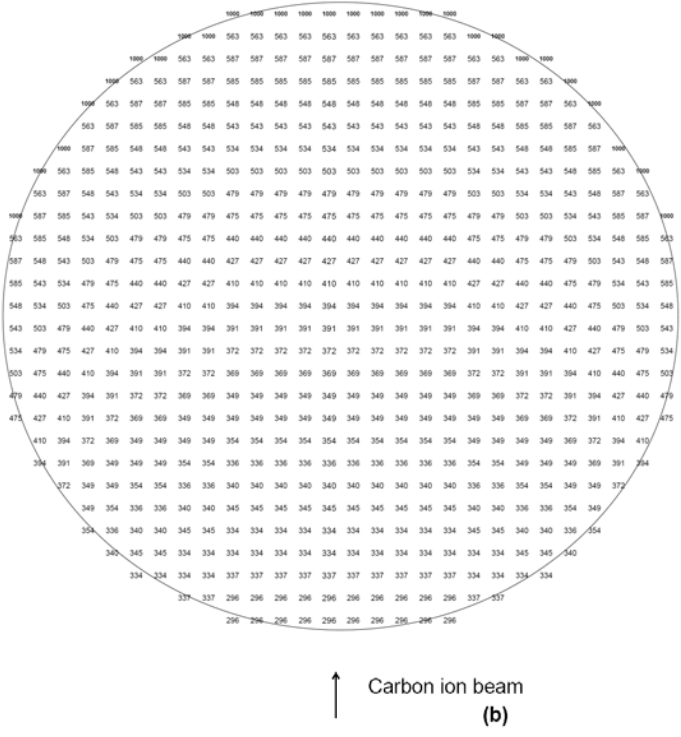

*Figure 7* *With spots having 5.0 mm FWHM dimensions, a ±1.25% uniform dose distribution is obtained by delivering to a 6.2 cm radius spherical volume centred at 20 cm of depth a number of particles proportional to the figures reported here.*

The numbers of particles are normalized to 1000. In the proton case the first slice requires, in the same units, the addition of only 70 protons because the corresponding tumour volume is crossed by all the protons that have their Bragg peak located deeper in the target.

The carbon situation is different because there is a difference between the "physical dose", measured in Gy, and the "equivalent dose" expressed in Gye obtained by multiplying the physical dose by the effective local RBE (Relative-Biological Effectiveness). This point is illustrated in Fig. 8 which is taken from Ref. [23].

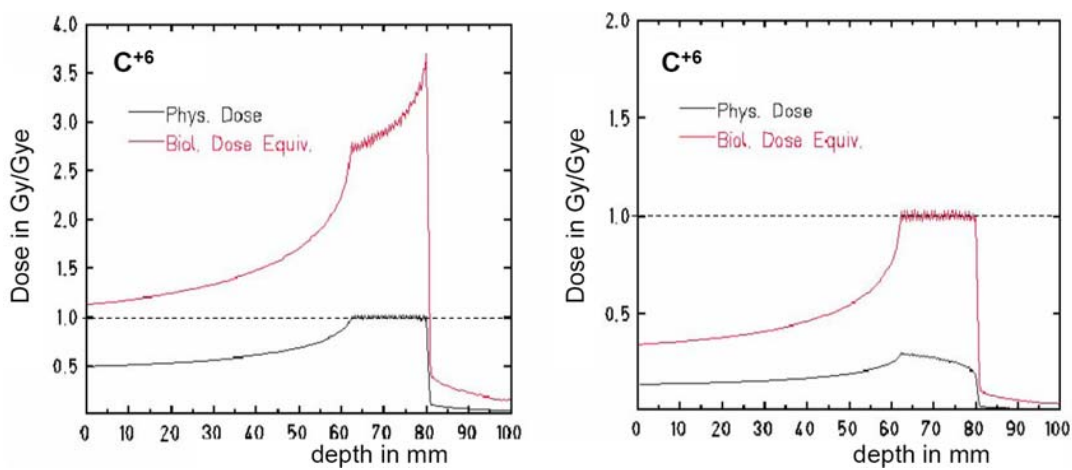

*Figure 8* Since the ions and the fragments reaching the distal part of the target have higher RBE, a constant physical dose of 1 Gy produces a biological equivalent dose which increases with the depth (left figure). To obtain a flat dose of 1 Gye, the physical dose has to be about three times smaller and decrease with the depth (right figure) [23].

The RBE has to be calculated with a model which mixes physics with radiobiology since in each point the radiation field is composed of both integer carbon ions and their upstream-produced fragments. The typical distributions of Fig. 8 have been obtained with the GSI model called LEM (Local Energy-deposition Model), which uses as input the response of the cells to X rays and computes the value of RBE as a function of the velocity and the type of the nuclear species involved. Since in the carbon case to obtain a 'flat' biological equivalent dose the physical dose has to decrease with the depth, the distal deliveries have to deposit less dose in the traversed tissues and the closer spots need more particles.

The two distributions of Fig. 7 can be obtained in a single painting but, for the reasons discussed in Section 3, many paintings can be essential in the treatment of moving organs. To define how many paintings are needed one can argue that in each point of the target a single spot never contributes more than 40% to the local dose, because of the contributions of the surrounding spots. The request of 12 *equal* deliveries is then an easily justifiable goal because the accidental absence of a spot would cause a local under dosage not larger than 3.3 % (i.e. 0.4 /12). It has to be noted that, at least in principle, such a miss can be corrected during the next paintings.

Fig. 9 shows how many 'visits' of the same voxel by the proton and carbon ion spots are needed to satisfy the stated above 3.3% condition.

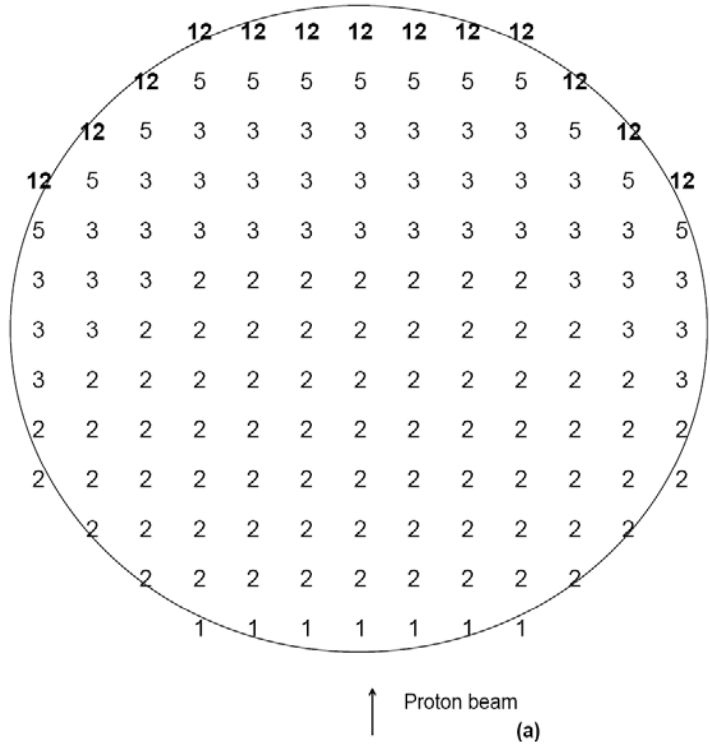

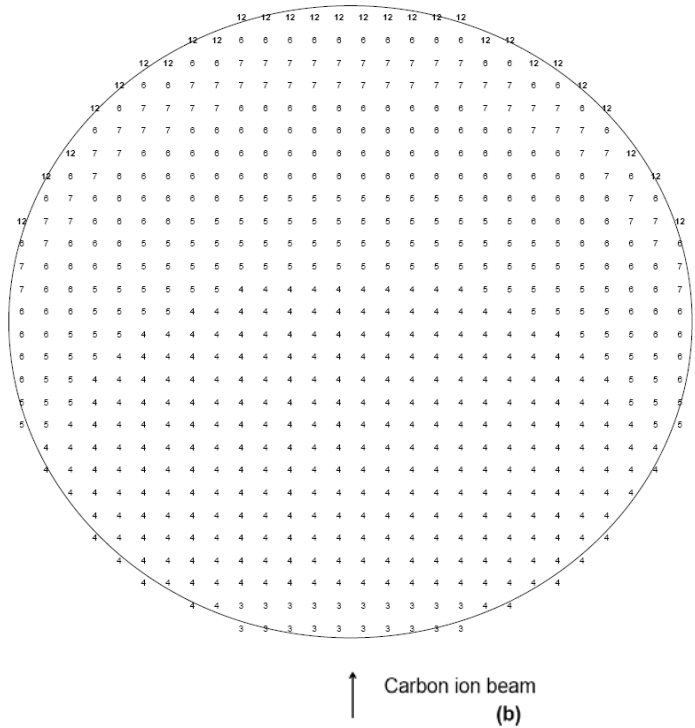

*Figure 9* *Number of 'visits' needed to obtain flat equivalent dose distributions with protons and carbon ions with the condition that any missing 'visit' does not reduce the dose by more than 3.3%.*

In three dimensions the total number of proton (carbon) deliveries is about 5,200 (79,800) so that with a 200 Hz (400 Hz) repetition rate, to homogeneously deliver the dose, when taking into account the dead times due the lowering of the magnetic fields in the delivery magnets during depth scanning, one needs 30s (220 s). It is seen that the proton and carbon cyclinacs described in Table 2 allow to give the 'standard' 2 Gye per minute and per litre if the maximum number of hadrons per linac pulse are $1.7\ 10^7$ and $1.5\ 10^5$ respectively as it can be evaluated from R2 in Table 3. As shown in Table 3 (row R3), the maximum number of hadrons corresponds to an average current not larger than 0.031enA, where the 'e' recalls that each carbon ion has charge Q=6.

*Table 3. Main parameters of the cyclinacs based on the two linacs of Table 2 in the 'many paintings' scenario of Figs. 7 and 9.*

| **Accelerated particles** | $p^{+1}$ | $C^{+6}$ | |
|---|---|---|---|
| Repetition rate [Hz] | 200 | 400 | R1 |
| Max. number of particles per pulse $N_m$ [for 2 Gy $L^{-1}$ $min^{-1}$ for ptotons and 4 Gye $L^{-1}$ $min^{-1}$ for carbon ions] | $3.3\ 10^7$ | $8.2\ 10^4$ | R2 |
| Max. current from the linac ($1.6\ 10^{-19}$xQxR1xR2) [enA] | 1.06 | 0.031 | R3 |
| Effective duration (FWHM) of a modulator power pulse [μs] | 3.2 | 3.2 | R4 |
| Duration of the RF pulse width [μs] | 1.5 | 1.5 | R5 |
| Cyclotron maximum current with standard DC sources [eμA] | 500 | 0.2-0.3 | R6 |
| Duty cycle of the HV power pulses (R1xR4) | $6.4\ 10^{-4}$ | $1.3\ 10^{-3}$ | R7 |
| HV wall plug power for the RF system (Table2: R16xR7) [kW] | 87 | 234 | R8 |
| Duty cycle of the RF pulses (R1xR5) | $3.0\ 10^{-4}$ | $6.0\ 10^{-4}$ | R9 |
| Linac computed phase acceptance for a continuous beam | 0.12 | 0.12 | R10 |
| Correction to R10 for possible linac misalignments | 0.80 | 0.80 | R11 |
| Cyclotron horiz. normalized transv. emittance $\varepsilon_{nH}$ [π mm mrad] | 2.0 | 5.0 | R12 |
| Cyclotron vert. normalized transv. emittance $\varepsilon_{nV}$ [π mm mrad] | 1.0 | 4.0 | R13 |
| Normalized transverse acceptance in both planes [π mm mrad] | 1.0 | 3.0 | R14 |
| Acceptance in transverse plane (from R12, R13 and R14) | 0.60 | 0.50 | R15 |
| Overall acceptance (R9xR10xR11xR15) | $1.7\ 10^{-5}$ | $2.9\ 10^{-5}$ | R16 |
| Cyclotron current needed for hadrontherapy (R3:R16) [eμA] | 62.4 | 1.08 | R17 |

To reduce the average high voltage wall plug power (R8) and therefore the running costs, the power pulse has been made relatively short: 3.2 μs (R4). This is a minimum because in a linac the klystrons are powered by 'modulators' which produce microsecond pulses at high voltages (100-150 kV) and medium currents (70 -150 A) having a rise-time and a fall-time of the order of 1 μs. With a usable 'flat top' of 1.5 μs (R5) the minimum effective length of the high voltage power pulse is 3.2 μs (R4). The high voltage duty cycle (R7) determines the average wall plug power consumed by the RF system (R8) which contains the 42% klystron efficiency (R15 of Table 2). Globally the $C^{+6}$ linac requires twice the high voltage power of the proton linac (R8) because of the larger aperture (R6 of Table 2) and the higher repetition rate (R1 of Table 3).

A non-negligible power has to be added to produce the klystron focusing fields. For instance, the TH 2157 klystron by Thales needs 7 kW for beam focusing and 2 kW for auxiliaries with a total of 9 kW when used in a classical line type modulator and with a 3 coil focal magnet. With 22 (18) klystrons for the proton and carbon ion linacs, the corresponding average electrical power for this scheme amounts to about 200 kW (160 kW). However, with the development of a mono coil focusing system and the use of only solid-state modulators, that have no additional heating power requirements, the total auxiliary electrical power can be reduced to 5.5 kW per klystron modulator. Additionally, as presented in Section 7, the use of the 7.5 MW klystron with fast ferrite transformer phase and amplitude correction [24] being done at the high level RF output leads to a 10 klystron system, since each klystron output RF power is split between two accelerating modules. In this way, the total additional electrical power for the same proton or carbon ion linear accelerator is reduced to 55 kW. This could be reduced even further if 3 GHz klystrons with permanent magnet focusing were developed to bring more power savings.

The longitudinal acceptance of the linac around the synchronous phase (R10, Table 2) was computed for a continuous input beam with computer codes provided by K. Crandall [25] because the linac RF period (0.33 ns, corresponding to 3 GHz) is definitely smaller than the duration (2-4 ns) of a cyclotron pulse.

An analysis of the effect of the misalignments of the PMQs indicates that the overall acceptance can be reduced at maximum by about 20%, considering misalignments with a flat distribution within 100 μm (R11).

The transverse emittances of the cyclotron beam enter in the overall acceptance of the linac. In Table 3 the emittance of the proton beams extracted from typical proton cyclotrons have been used (R12 and R13). For the carbon linac the characteristics of the LNS cyclotron have been extrapolated from the published data [22] and comparisons with smaller superconducting cyclotrons.

The rows R14 and R15 give the computed linac acceptance and the fraction of the cyclotron beam which is transversally accepted. All the above factors multiply in the overall acceptances of the two linacs (R16) which are in the range 2-3 $10^{-5}$ (R16), as anticipated at the beginning of the last Section, and define the needed cyclotron currents (R17) of about 60 and 1 eμA, respectively.

By comparing the figures in R16 and R6, it can be concluded that the proton currents are about 15 times larger than what is needed, while the $C^{+6}$ current extracted from the SC cyclotron is not sufficient by a factor 5. In fact the existing "Supernanogun" ECR sources by Pantechnik (France) [26] produce a 3 eμA continous beam of $C^{+6}$ ions that corresponds to about 0.2-0.3 eμA with a 7-10% cyclotron acceptance. A factor of 5 has thus to be gained. According to the constructor a "Supernanogun" source could be pulsed at 400 Hz so to multiply the $C^{+6}$ current by a factor 3-4.

Recently the Company DREEBIT GmbH (Germany) has developed a superconducting Electron Beam Ion Source (EBIS) which is better suited [27]. In fact this source provides ion pulses with pulses varying between 1 μs and 100 μs. At a 400 Hz repetition rate, each pulse contains up to 4 $10^8$ $C^{+6}$ ions, more than enough for the cyclinac of Table 3 [28].

Moreover one could use two sources in parallel, since the cyclotron acceptance is large enough. Two other alternatives are the superconducting ECR source "Hypernanogun" by Pantechnik (which on paper produces much more than the needed 1 eμA, but measurements have as not yet been performed) and the last generation of ECR Ion Sources (ECRIS) [27].

## 6. CYCLINACS FOR PROTONTHERAPY: "IDRA"

The integration of proton cyclinacs in a fully-fledged diagnostic and therapy centre has been studied in detail under the name 'IDRA'. This stands for *Institute for Diagnostics and RAdiotherapy* and is a fitting name since in medieval latin the name Idra ('Hydra' in English) designated a mythologic creature with many very vital 'heads' [29], just as a high-current cyclotron.

Three projects have been pursued. The first one has been described in Section 5 (Tables 2 and 3) and is based on a 30 MeV cyclotron.

The second project has at its heart the 70 MeV cyclotrons built either by IBA [30] or by the Joint Institute for Nuclear Research (JINR, Dubna) [31]. The design of the linac is very similar with respect to the one employing a 30 MeV cyclotron and the main difference is that 14 modules are needed to reach 240 MeV for a 14.6 m total length. The accelerating gradients in the tanks are in the range 18.2 – 18.5 MV/m and the PMQ gradients fall in the range 166 – 128 T/m.

In the third project the cyclotron accelerates protons to 15 MeV and the linac is composed of a DTL section of the CLUSTER type from 15 MeV to 67 MeV and a SCL section of the LIBO type from 67 MeV to 235 MeV. To improve the linac efficiency and to reduce the overall length, CLUSTER works at 1.5 GHz and LIBO at 3 GHz. CLUSTER is a novel IH transverse electric field structure running at high frequency [32]. A module is sketched in Fig. 10 and the main parameters are reported in Table 4. It is important to remark that 15 MeV cyclotrons are today the most common type of accelerators used in hospitals to produce radioisotopes for PET diagnostics (Fluorine 18 in particular).

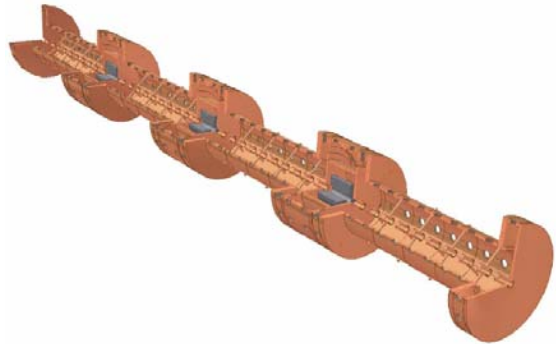

*Figure 10. A module of CLUSTER, the "Coupled-cavity Linac USing Transverse Electric Radial" field. The accelerating tank consists of a sequence of identical (constant β) accelerating units each one formed by an accelerating gap and by two half drift tubes. In CLUSTER the transverse electric field is brought horizontally along the axis by the copper drift tubes. The accelerated beam is focussed by PMQs.*

The need of high power efficiency in the low β range (0.05-0.3) leads to the choice of H-mode accelerating cavities, also called TE cavities because the electric field is naturally directed transversally with respect to the structure axis. H-mode cavities are drift tube cavities operating in the $H_{n1(0)}$ mode, where the index *n* is usually 1 (IH cavities; already existing) or 2 (CH cavities, under development). These cavities are very attractive because of the high shunt impedance for low β's due to the fact that the generally transverse electric field is made parallel to the axis and concentrated in the accelerating gaps by the metallic drift tubes. Moreover, they are π-mode structures, i.e. the RF accelerating field is phase shifted by 180° between successive gaps, a feature allowing higher average gradients which in the present case are further increased by the choice of a large frequency (1.5 GHz).

*Table 4. Main parameters of the linac which accelerates protons from 15 MeV to 235 MeV*

| **Type of linac** | **DTL CLUSTER** | **SCL LIBO** | |
|---|---|---|---|
| Frequency [MHz] | 1499 | 2998 | R1 |
| Input energy [MeV/u] | 15 | 67 | R2 |
| Output energy [MeV/u] | 67 | 235 | R3 |
| Number of accelerating cells per tank | 7 | 14 | R4 |
| Number of accelerating structures per module | 8 - 6 - 4 - 4 | 2 | R5 |
| Number of accelerating modules (for the SCL = n. of klystrons) | 4 | 16 | R6 |
| Diameter of the beam hole [mm] | 8 | 8 | R7 |
| Total length of the linac [m] | 6.2 | 12.8 | R8 |
| Number of PMQs [length = 30 mm; gradient = 126-170 T/m) | 19 | 32 | R9 |
| Maximum misalignment for the PMQs [mm] | ±0.1 | ±0.1 | R10 |
| Synchronous phase | -13° | -15° | R11 |
| Peak power per module with 10% losses in waveguides [MW] | 2.0 – 2.4 | 3.4 | R12 |
| Effective shunt impedance $ZT^2$ (inject.-extract.) [MΩ/m] | 109 - 81 | 43 - 70 | R13 |
| Average axial electric field (injection-extraction) [MV/m] | 14.7 | 16.7-19.0 | R14 |
| Number of klystrons (peak powers 20 MW and 7.5 MW) | 1 | 16 | R15 |
| Total peak RF power for all the klystrons [MW] | 9.0 | 54 | R16 |
| Repetition rate of the proton pulses [Hz] | 200 | 200 | R17 |
| Pulse length [μs] | 5.0 | 5.0 | R18 |
| Duty cycle [%] | 0.1 | 0.1 | R19 |
| Power required by the linac (R16xR19) [kW] | 9.0 | 54 | R20 |
| Klystron efficiency | 0.42 | 0.42 | R21 |
| Min. HV wall plug power for the RF system (R20:R21) [kW] | 21 | 128 | R22 |

The rest of this Section is devoted to the first project, the one described in Tables 2 and 3 of Section 5, since this is the one which has been studied in more detail.

This design features:
- a 30 MeV high-current commercial proton cyclotron with several external beams,
- a high-gradient linac – essentially based on the LIBO prototype - which accelerates protons from 30 MeV to 235 MeV,
- fixed beams and/or rotating gantries for the treatment of deep seated tumours.

As schematically shown in Fig. 11, such an accelerator complex features on the same site:
- the production of - and the research on - radiopharmaceuticals for PET, SPECT and other new diagnostic and therapeutic medical applications,
- the treatment of deep seated tumours with proton beams of energy as large as 235 MeV (corresponding to a 33 cm range in water).

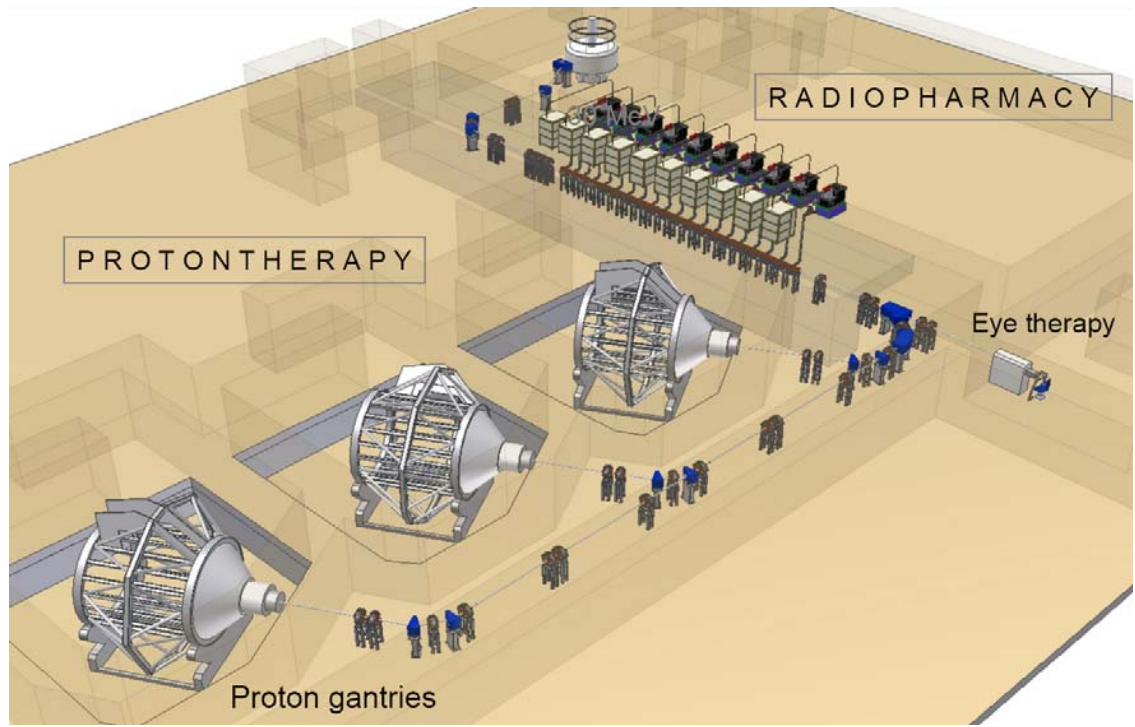

*Figure 11. A typical layout of IDRA features a 30 MeV cyclotron a linac of the LIBO type and three treatment rooms equipped with rotating gantries and a fixed beam line for the treatment of eye tumours.*

IDRA can be realized in successive phases. In Phase 1, with six accelerating modules, the protons would reach the 70 MeV needed for eye therapy. Adding 10 modules in Phase 2 protons could be accelerated to 170 MeV so to treat all paediatric and head and neck tumours. Phase 3 would see the completion of the layout of Fig. 11.

A standard delivery system would comprise a horizontal beam and two or more isocentric gantries working with the spot scanning technique developed at PSI. In fact, as discussed in Section 2, the therapy beam produced by a cyclinac is more suited to patient treatment with the many-paintings 'spot scanning' technique than the beams produced by cyclotrons and synchrotrons.

It has to be stressed that the research possibilities opened by a high-current 30 MeV cyclotron are much wider than the ones made available by a 'standard' hospital cyclotron producing 15-18 MeV protons. The first and most obvious is the production of radioisotopes for diagnostics and therapy purposes, as for instance Alfa immunotherapy [33].

The 500 microampere 30 MeV proton beam can also produce a large flux of fast, epithermal and thermal neutrons by using a suitable moderator. This opens the way to at least three very promising medical applications.

The thermal or epithermal neutron field can be used for *Boron Neutron Capture Therapy* (BNCT) [34] and *Boron Neutron Capture Synovectomy* (BNCS) [35]. These techniques are still experimental, but they promise the cure of some radio resistant tumours (BNCT) and the palliation of rheumatoid arthritis (BNCS). Intense neutron fluxes can be used to apply the new technique of radioisotope production called "*Adiabatic Resonance Crossing*" (ARC) [36].

In summary, IDRA opens the way to protontherapy, radioisotope production and other potential interesting medical applications and is a physical and cultural space in which radiation oncologists, nuclear medical doctors, nuclear chemists, medical physicists, physicists and engineers can work together towards the common goal of diagnosing and curing solid tumours, their metastases and other diseases.

## 7. THE LINAC OF IDRA

As far as the linac is concerned, modifications have been introduced to the prototype design of Fig. 5, which is made of four tanks fed through a single bridge coupler. If the structure of the prototype were maintained, the final module would be more than two metres long and therefore difficult to handle and braze. Thus each one of the 20 accelerating modules of IDRA is made of two tanks powered through a bridge coupler, as shown in Fig. 12 (see also Fig. 6).

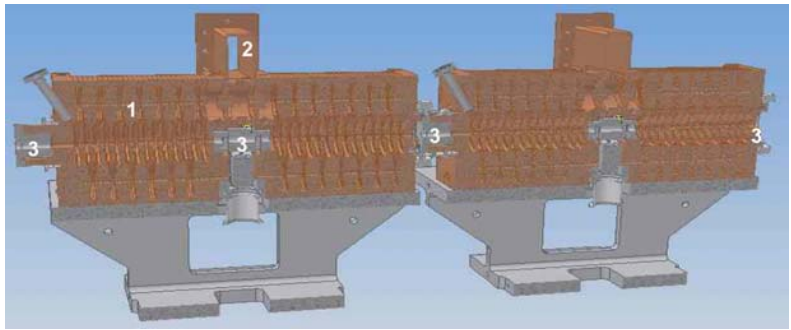

*Figure 12.* *The first two modules accelerate protons from 30 to 35 MeV and from 35 to 40 MeV: 1. accelerating 'tank', 2. bridge coupler, 3. locations of the permanent quadrupole magnets (PMQs).*

The first of the 22 accelerating modules is 55 cm long. To maintain the synchronism between the RF phase and the proton velocity, which increases from about 25% to about 60% of the speed of light the last module is 110 cm long.

As shown in Fig. 12, the bridge couplers and the end cells (which are thick copper plates) house the permanent magnet quadrupoles (PMQs) needed to keep the beam focused during the acceleration. The twenty modules are water cooled and the temperature of the water (around 27°C) is used for the fine tuning of the resonance frequency.

The average accelerating field of each module is such that the peak power is about 2.5 MW for each one of them (R11, Table 2). The power is provided by 2.998 GHz klystrons, which are powered by a 'modulator'. Modulators of the needed characteristics are standard items which can be provided by many companies.

Twenty modulators and twenty klystrons require a large investment and a reduction in the number of the units implies both a reduction of the installation space and a lowering of the cost. Moreover a medical facility has to run continuously and reliably and the possible failures of the large number of klystrons required by IDRA may be a problem.

A proposed innovative approach is based on novel devices developed at CERN for LEP (the Large Electron Positron Collider) to work at 352MHz and re-designed for 3 GHz operation in IDRA. The basic principle of operation of the main component, a fast ferrite transformer called IQM [37], is schematically shown in Fig. 13.

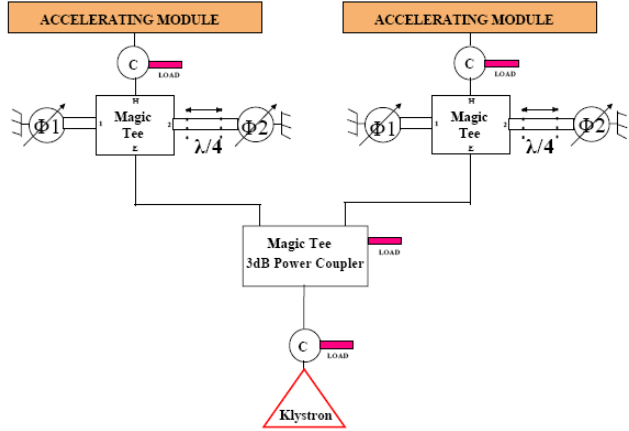

*Figure 13. The phase-shifting and attenuation methods using reflective type IQM modulators based on fast ferrites are shown for one klystron [38] of IDRA.*

The klystron output is connected, downstream of an appropriate "load" needed to absorb the reflected power, to a 3dB Magic Tee power splitter. This component splits equally the power between the two accelerating modules.

As shown in the Fig. 13, a fast ferrite transformer, called IQM, is placed between the power splitter output and each accelerating module input. An IQM enables dynamic (~3 ms) for precise pulse to pulse amplitude adjustments of the high power RF.

The functioning principle is simple. The two RF waves from the output of a Magic Tee are individually phase shifted and then recombined at the single output RF port. When the fast ferrite transformer phase shifters are DC biased so that the individual RF signals are in anti-phase then the amplitude of the output signal changes. When the signals are in phase, the output phase can be adjusted. By a combination of these two modes both phase and amplitude can be changed simultaneously. However for cyclinac applications the power adjustment is sufficient.

As anticipated in Section 3, the number of accelerating modules and the structure of the FODO focussing channel have been chosen in such a way that the outgoing beam energy can be adjusted continuously. The depth dose distributions of Fig. 14 have been computed by tracking the proton trajectories inside the linac [38]. The figure shows that, by having a variable number of active klystrons, the penetration depth can be varied between 20 mm and 330 mm and this can be done in only 1 millisecond because only the low-power signals of the drivers have to be switched off and on. Intermediate energies, and thus ranges, can be obtained by varying the power transmitted by the last active IQM fast ferrite transformer device.

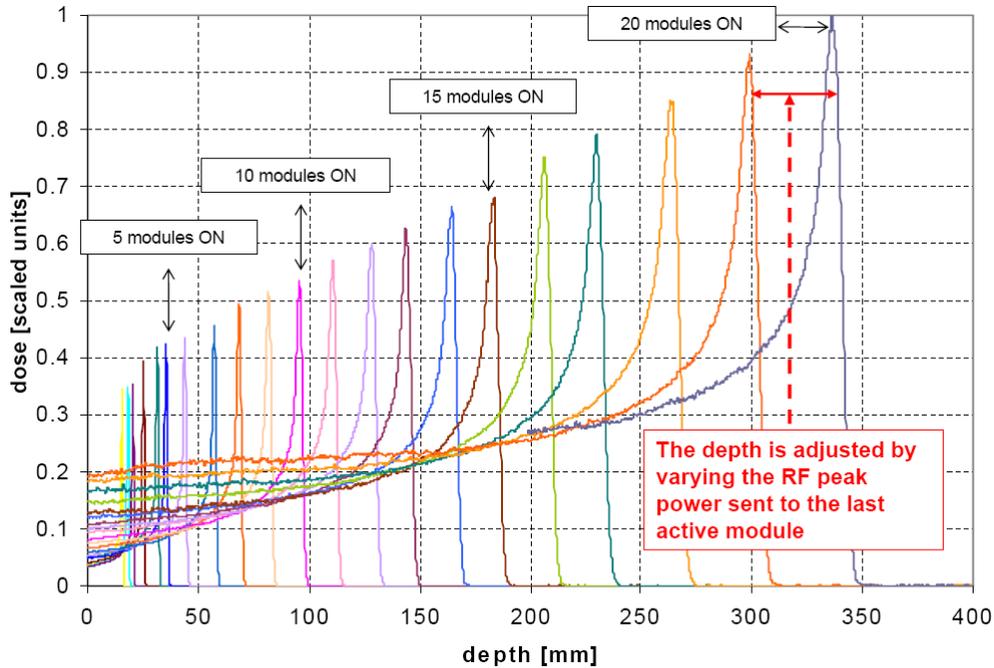

*Figure 14. Proton depth dose distribution when the number of the active accelerating modules is varied. To avoid super positions a different normalization is used for each curve.*

The variation of the power level injected in the accelerating modules cause a small variation of the beam energy spread. However dynamics calculation show that the consequent range spread is definitely smaller than the natural spread due to straggling.

In the treatment rooms of Fig. 11 the dose is delivered by spot scanning with at least 12 paintings (Table 3). During a patient treatment, spot after spot two quantities are adjusted: the energy E of the protons in the range 30 MeV ≤ E ≤ 235 MeV - as discussed above - and the number N of protons is adjusted by acting on the beam elements upstream and downstream of the cyclotron. These tunings, covering collectively the range $N_m/50 \leq N \leq N_m$, are performed in 3 ms so to obtain the dose distribution planned with the TPS. The charge can be predetermined with 3% accuracy.

## 8. A CYCLINAC FOR CARBON ION THERAPY: "CABOTO"

A suitably modified version of the $C^{+6}$ cyclinac described in Table 2 and 3 has been designed for the ion-therapy centre proposed by LNS of INFN. In this project, the superconducting cyclotron SCENT (*Superconducting Cyclotron for for Exotic Nuclei and Therapy*) accelerates $C^{+6}$ ions to 300 MeV/u [39] instead than to 250 MeV/u initially chosen [22]. Since the end of 2006, the Belgian company Ion Beam Applications (IBA) proposes commercially SCENT at 300 MeV/u which accelerates both hydrogen molecules ($H_2^+$), extracted as protons by a thin foil, and carbon ions ($C^{+6}$) usable for treating shallow tumours.

As presented in Fig. 15, the results obtained in the years 1995-2002 by H. Tsujii et al at HIMAC in Japan [13] show that the gain in depth from 12 cm to 17 cm when passing from 250 MeV/u to 300 MeV/u allows the treatment of a larger fraction of patients [40].

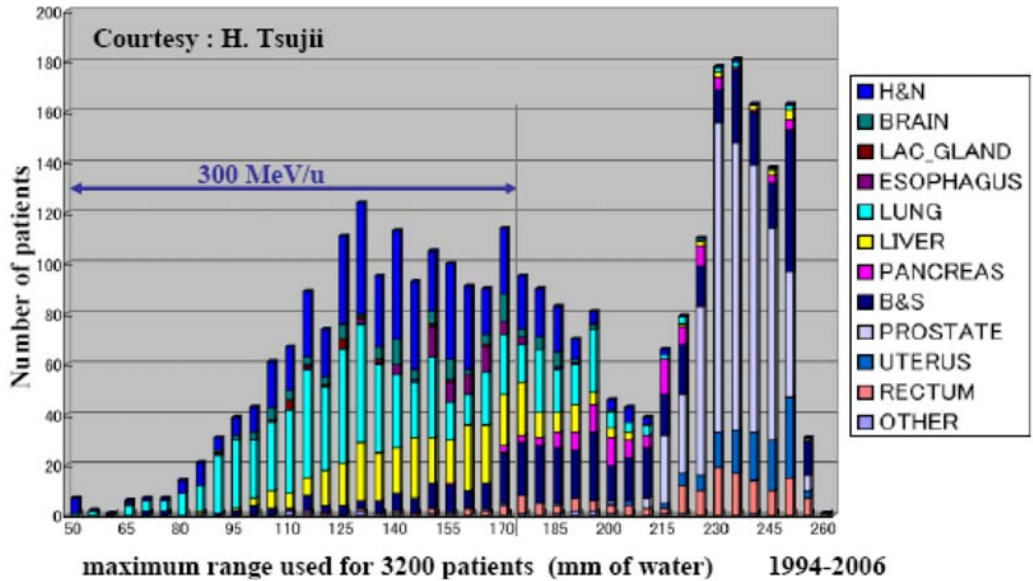

*Figure 15.* On the vertical axis the number of patients is plotted while the horizontal axis represents the water equivalent depth of the maximum range used for each patient. The 2000 HIMAC patients had the tumours indicated in the inset [40].

The figure shows that, according to the HIMAC experience, the 17 cm range of 300 MeV/u carbon ions is sufficient to treat 85% of the head and neck tumours, 80% of bone and soft tissues sarcomas, ≤ 20% pancreas and prostate carcinomas and ≤ 3% of other tumours.

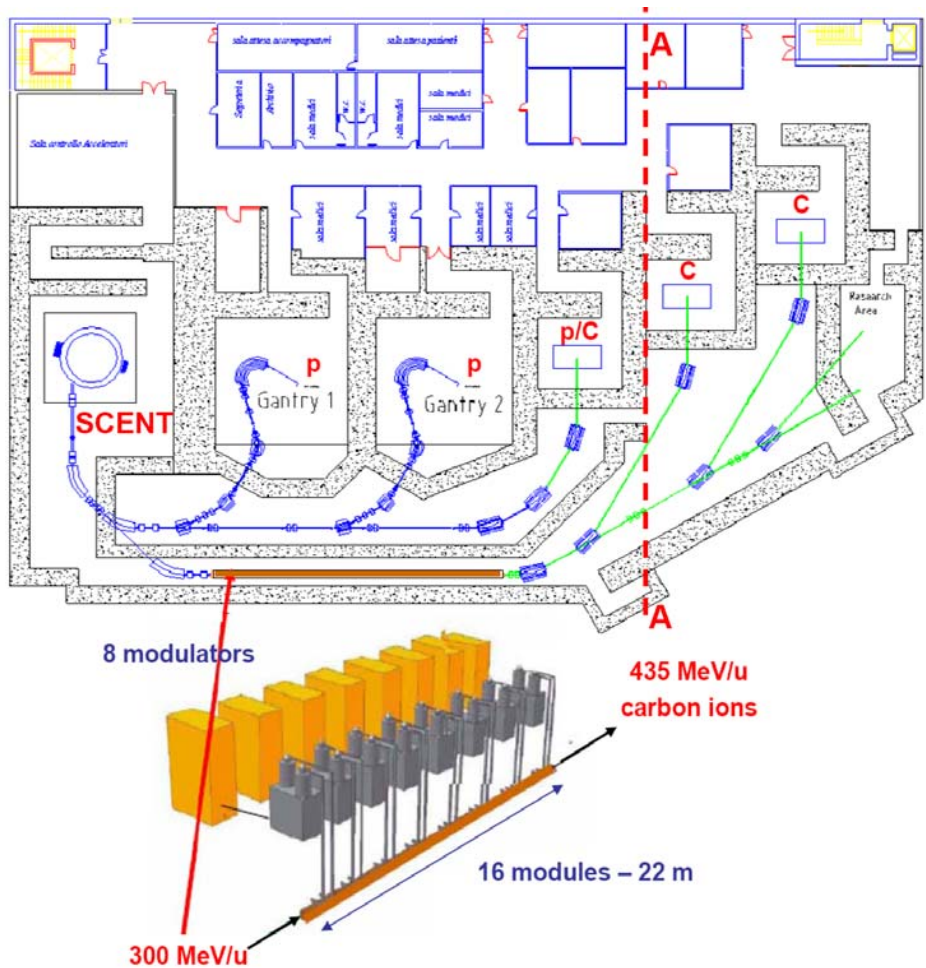

*Figure 16. The hadrontherapy centre designed by the Catania group is the one at the left of the AA line. The installation of the 8 unit linac (16 modules) and the addition of the building at the right of the AA line will allow to reach with carbon ions a depth equivalent to 32 cm of water.*

The layout of the proposed centre is reported in Fig. 16 and features three rooms for protontherapy, two of them featuring proton gantries. In the treatment room indicated as p/C superficial tumours can be treated also with 300 MeV/u carbon ions, which have a range of 17 cm in water. To reach a depth of more than 30 cm of water, it has been proposed to add a linear carbon ion accelerator – named *CABOTO* (*CArbon BOoster for Therapy in Oncology*). As shown in Fig. 16, this linac has 16 accelerating modules of the LIBO type and allows to reach energies of 435 MeV/u .

The first accelerating module, made of two 'tanks', is represented in Fig. 17. It has to be noted that the brazing of these modules requires larger ovens than the ones needed for the production of the modules of the "proton cyclinac" discussed in the previous Section. On the other hand, the fact that the composing elements, the so-called "half-cells", are thicker brings a mechanical simplification. The total length of the linac of Fig. 16 is 22 metres.

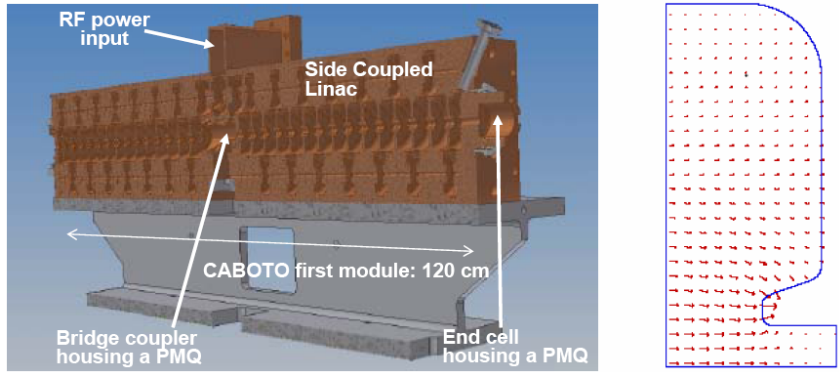

*Figure 17.* *Each one of the two tanks of a CABOTO module is made of 15 accelerating cells (left figure). In between two contiguous tanks a Permanent Magnet Quadrupole focuses the ion beam. The figure on the right shows the map of the electric field in an accelerating cell.*

Taking into account losses of the order of 10%, each of the 16 accelerating modules requires 4.2 MW. In the scheme shown in Fig. 16, 16 klystrons identical to the ones used for IDRA and producing 7.5 MW, are powered two-by-two by 8 modulators.

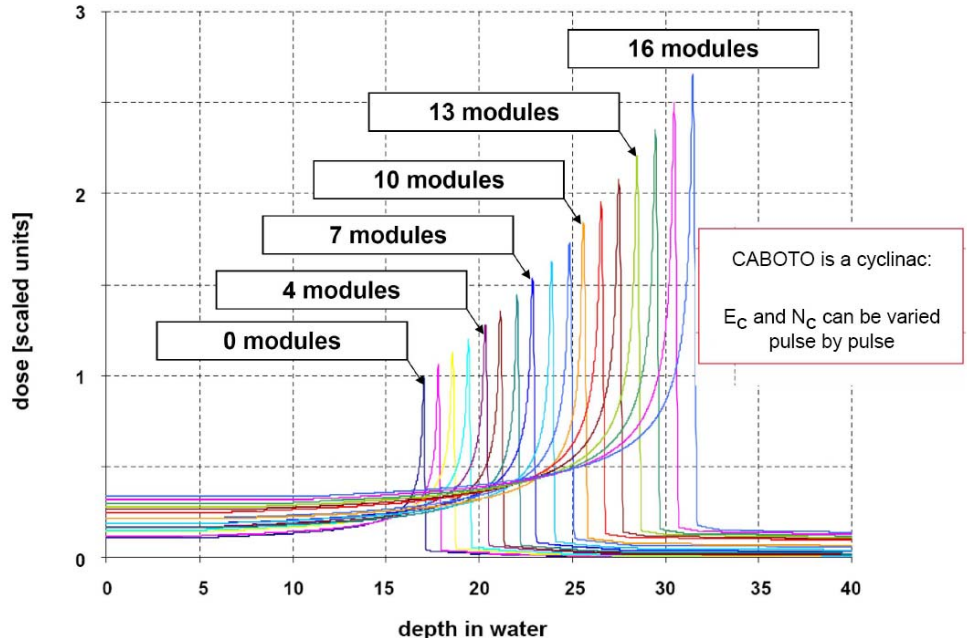

*Figure 18.* *The Bragg curves obtained by varying the number of active modules are shown by rescaling the peaks so to avoid super positions.*

Independently from the powering system, as in the case of IDRA, the energy of the carbon beam can be varied in about 3 milliseconds to give the depth-dose distributions presented in Fig. 18. This feature – as already pointed out – leads also in the case of carbon ions to very important potentialities to respond to the challenges hadrontherapy will face in the next future, as presented in Section 2.

\*\*\*

The cyclinac for carbon ions (CABOTO) [41] and the CLUSTER accelerating structure [42] are covered by patents. A patent application has been deposited for the cyclinac for protons (IDRA) [43].

**Acknowledgments**

We are grateful to Sonia Allegretti for her contributions to the early mechanical designs of IDRA.

The financial support of the *Monzino Foundation* (Milano), the *Price Foundation* (Geneva) and the *Associazione per lo Sviluppo del Piemonte* (Torino) are gratefully acknowledged.